# Frictional magnetodrag between spatially separated two-dimensional electron gases mediated by virtual phonon exchange

Samvel Michael Badalyan

*Department of Radiophysics, Yerevan State University, 375025 Yerevan, Armenia*

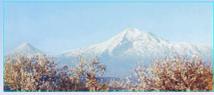
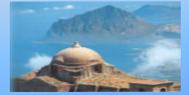

## Theoretical model

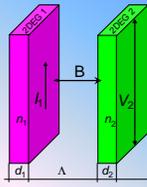

- The model consists of two spatially separated 2DEGs with densities $n_1$ and $n_2$ and layer thicknesses $d_1$ and $d_2$. Both electron sheets are exposed to the perpendicular magnetic field $B$.
- Frictional magnetodrag manifests itself when a current driven along layer 1 induces, via momentum transfer, a drag voltage in the layer 2.

Direct measures for the interlayer e-e interaction in this double layer system are the transresistance and the drag scattering rate which are defined as

$$\rho_{Drag} = \frac{V_2}{I_1} \text{ or } \frac{1}{\tau_{Drag}} = \frac{E_2}{\tau_1 E_1}.$$

## e-e interaction mediated by phonons

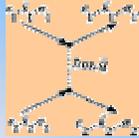

From the Fermi's Golden Rule

$$W^p_{1,2\to1',2'} = \frac{2\pi}{\hbar}|T^p_{1,2\to1',2'}|^2 \delta(\varepsilon_1 + \varepsilon_2 - \varepsilon_{1'} - \varepsilon_{2'})$$

$T$ is the transition matrix element and in the second order of the PT is given by this diagram.

Each solid dot corresponds to the electron phonon vertex part $\Gamma$ in this diagram.

$$\Gamma^p_{ll'}(q) = \sqrt{B^p(q)}\delta_{l,l'+q_z} Q_{ll'}(t)\int dz \rho(z) \exp(iq_z z),$$
$$Q_{ll'}(t) = \left(\frac{l!}{l'!}\right)^{1/2} e^{-t^2/2} t^{l'-l} L_l^{l'-l}, \ t^2 = \frac{q_\perp^2 l_B^2}{2},$$
$$B^p(q) = q^{-1}B_0^{PA}, qB_0^{DA}, q^{-2}B_0^{PO} \text{ for PA, DA, PO}.$$

The dashed line corresponds to the phonon propagator $D(q,\omega)$, where $\omega_q$ and $\tau_q$ are the phonon energy and lifetime.

$$D^p(q,\omega) = \frac{2\hbar^{-1}\omega_q^p}{(\omega + i/2\tau_q)^2 - \omega_q^2}$$

## Drag scattering rate

The drag scattering rate, which can be derived for instance from either the **Boltzmann linearized equation** or **Kubo linear response formula**, is given by

$$\frac{1}{\tau^p_{Drag}} = \frac{\hbar^2}{4m^*Tn_\perp L^2} \sum_{1,2,1',2'} \int dq \, q^2_\perp \frac{W^p_{1,2\to1',2'}}{2\pi|\varepsilon(q_\perp,\omega)|^2} \frac{\text{Im}\,\chi_1(q_\perp,\omega)\text{Im}\,\chi_2(q_\perp,\omega)}{\sinh^{-2}(\hbar\omega/2T)} \quad (1)$$

$q_\perp$ is the transferred momentum, $\varepsilon(q_\perp,\omega)$ is the screening function in the RPA, $\chi(q_\perp,\omega)$ is the polarization function, $W_{1,2\to1',2'}$ is the transition probability of two electrons from the states $|1\rangle$, $|2\rangle$ into $|1'\rangle$, $|2'\rangle$. To avoid unphysical jumps and singularities we use a Gaussian density of states

$$g_l(\varepsilon) = \frac{\sqrt{2/\pi}}{2\pi l_B^2 \Gamma_0} \exp\left(-2\frac{(\varepsilon-\varepsilon_l)^2}{\Gamma_0^2}\right), \ \varepsilon_l = (l+1/2)\hbar\omega_B, \ \Gamma_0 = \frac{2}{\pi}\hbar\omega_B\frac{\hbar}{\tau}$$

$\tau$ is the transport time related to the mobility, $l_B$ and $\omega_B$ are the magnetic length and the cyclotron frequency. The chemical potential is determined from

$$n = \sum_l \int_0^\infty d\varepsilon f_F(\varepsilon-\mu) g_l(\varepsilon), \text{ with } f_F$$

being the Fermi function.
We calculate polarization function $\chi(q_\perp,\omega)$ from Lindhard formula by using electron Green functions with imaginary part corresponding to the Gaussian and delta-function density of states.

## Magnetodrag due to virtual phonons

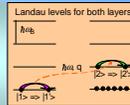

In many experiments $\hbar\omega_B \gg T, \hbar/\tau$ and electron transitions $|1\rangle\to|1'\rangle$ and $|2\rangle\to|2'\rangle$ in each layer are in the **same** partially filled outermost Landau levels. Therefore the drag is realized due to exchange of phonons **with zero energy and finite momenta**. The phonon propagator $D(q,\omega)$ is never on the "mass surface" since $\omega=0$ and $\omega_q \neq 0$, magnetodrag is only mediated **by virtual phonons**. Magnetodrag can be due to exchange of **virtual acoustic** and **optic** phonons. Two factors determine their relative contributions: the **optic** phonon propagator $D(q,\omega)$ is smaller at $\omega=0$ while the vertex part $\Gamma$ is larger than that of **acoustic** phonons.

## Magnetic field, temperature, and interlayer spacing dependence of the drag scattering rate due to piezoelectric acoustic phonons

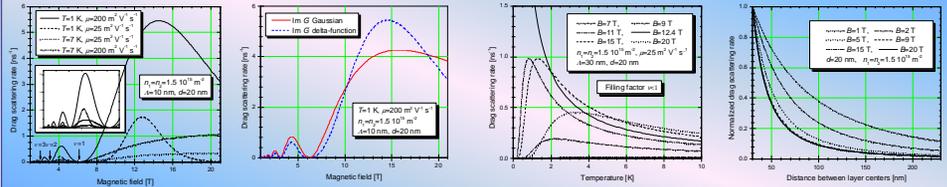

**Fig. 1.** Drag scattering rate as a function of magnetic field at various temperatures without the screening effects, i.e. $\varepsilon(q_\perp,\omega)=1$ in Eq. (1). In calculations of the functions $\chi(q_\perp,\omega)$ we use the electron Green functions corresponding to the delta-function density of states (the left figure) and the result is compared with the calculations for the Gaussian density of states (the right figure). Both approximations are in good agreement. The drag rate exhibits pronounced Shubnikov-de Haas oscillations.

**Fig. 2.** Temperature $T$ dependence of the drag rate for various magnetic fields. At high $T$, the drag rate decreases as $T^{-1}$ while at low $T$ it increases as a power function. At half filling the drag rate diverges at zero $T$.

**Fig. 3.** Drag rate as a function of interlayer separation without the screening effects for various magnetic fields normalized to its value at 30 nm center-to-center separation of two 20 nm wide 2DEGs.

## Double peak structure due to the screening effects

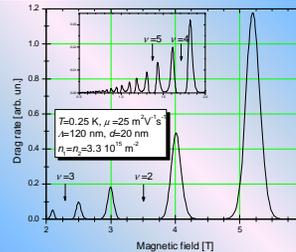

**Fig. 4.** Drag scattering rate as a function of magnetic field at various temperatures with the screening effects by $\varepsilon(q_\perp,\omega)$ in Eq. (1) and for the completely spin degenerated case. The drag rate shows a double-peaked structure as a function of filling factor, firstly revealed in a recent theory for the Coulomb drag by Bønsager et al [1] and observed experimentally by Rubel et al [2].

## Drag rate in an asymmetric double quantum well system

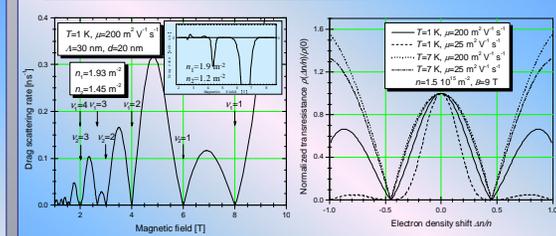

**Fig. 5.** Drag scattering rate at mismatched densities as a function of the magnetic field without the screening effects. Inset shows the drag rate on assumption that the drag is negative when the outermost Landau level of one layer is more than half filled while the other is less than half filled [3] with the screening effects included.

**Fig. 6.** Normalized dependence of the transresistance on shift of electron density from one layer to the other for various temperatures and mobilities. The total electron density remains unchanged and the average Landau level filling factor does not correspond to the half filling factor.

## Conclusions

- We have calculated the magnetodrag scattering rate between spatially separated 2D electron layers due to virtual phonon exchange with and without screening effects.
- The drag rate as a function of magnetic field shows the Shubnikov-de Haas oscillations without the screening effects. The screening effects result to the double-peaked structure, firstly revealed for the Coulomb magnetodrag by Bønsager et al [1] and observed by Rubel et al [2,3].
- The drag rate as a function of $T$ has a peak without the screening effects. At small $T$ it increases as a power-law function and at high $T$ it decreases as $T^{-1}$. At half filling the drag rate diverges at $T=0$. The screening eliminates this divergence and results to its monotonic increase in $T$.

## Acknowledgments

This research was partially supported by the Ministry of Science and Education of Armenia. The author acknowledges the support and hospitality at MPI SSR (Germany) and ICPT (Italy) where part of this work was done and thanks the members of Prof. K. von Klitzing's research group for useful discussions.